\documentclass[conference]{IEEEtran}
\IEEEoverridecommandlockouts
\usepackage{cite}
\usepackage{amsmath,amssymb,amsfonts}
\usepackage{algorithmic}
\usepackage{graphicx}
\usepackage{textcomp}
\usepackage{xcolor}
\usepackage{booktabs}
\def\BibTeX{{\rm B\kern-.05em{\sc i\kern-.025em b}\kern-.08em
    T\kern-.1667em\lower.7ex\hbox{E}\kern-.125emX}}
\begin{document}

\title{Missing-Data Handling in Sparse Outdoor BLE Positioning: The Role of Temporal Processing\\
\thanks{This work was supported by JSPS KAKENHI Grant Number 26H02547.}
}

\author{\IEEEauthorblockN{1\textsuperscript{st} Yu Sun}
\IEEEauthorblockA{\textit{Graduate School of System Informatics} \\
\textit{Kobe University}\\
Kobe, Japan \\
yu.sun@fine.cs.kobe-u.ac.jp}
\and
\IEEEauthorblockN{2\textsuperscript{nd} Zhiyi Zhu}
\IEEEauthorblockA{\textit{Graduate School of System Informatics} \\
\textit{Kobe University}\\
Kobe, Japan \\
syu@port.kobe-u.ac.jp}
\and
\IEEEauthorblockN{3\textsuperscript{rd} Patrick Finnerty}
\IEEEauthorblockA{\textit{Graduate School of System Informatics} \\
\textit{Kobe University}\\
Kobe, Japan \\
finnerty.patrick@boar.kobe-u.ac.jp}
\and
\IEEEauthorblockN{4\textsuperscript{th} Takenao Ohkawa}
\IEEEauthorblockA{\textit{Graduate School of System Informatics} \\
\textit{Kobe University}\\
Kobe, Japan \\
ohkawa@kobe-u.ac.jp}
\and
\IEEEauthorblockN{5\textsuperscript{th} Kenji Oyama}
\IEEEauthorblockA{\textit{Graduate School of Agricultural Science} \\
\textit{Kobe University}\\
Kobe, Japan \\
oyama@kobe-u.ac.jp}
\and
\IEEEauthorblockN{6\textsuperscript{th} Chikara Ohta}
\IEEEauthorblockA{\textit{Graduate School of System Informatics} \\
\textit{Kobe University}\\
Kobe, Japan \\
ohta@port.kobe-u.ac.jp}
}

\maketitle

\begin{abstract}
Sparse Bluetooth Low Energy (BLE) reception leaves learning-based localization with incomplete instantaneous inputs. We examine how retaining and filtering past received signal strength indicator (RSSI) observations affect localization in a ten-day outdoor deployment with four gateways and sixteen animals. Random-forest and nearest-neighbour regressors serve as single-frame probes, with validation-based parameter selection and paired animal-cluster bootstrap 95\% confidence intervals. In an evaluation where GPS does not screen reception inputs, a four-second hold reduces random-forest error by 8.092 m [6.797, 9.476] compared with instantaneous inputs. Separately, input filtering improves on a 120-second hold by 0.810 m [0.381, 1.309]. With separately validation-selected output filters, the short-hold benefit remains 7.336 m [6.127, 8.568], while the input-filtering benefit over the 120-second hold is 0.121 m [$-0.317$, 0.584]. Representation ablations, elapsed-time state-space controls, and day-block analyses assess sensitivity to the evaluated configurations. Short histories provide substantial information in this deployment, and further input filtering should be evaluated together with downstream temporal processing.
\end{abstract}


\section{Introduction}
Low-cost outdoor localization must accommodate limited infrastructure and intermittent reception. A learning-based RSSI estimator maps a fixed-dimensional signal representation to position, but an instantaneous observation may contain only a small subset of the expected gateway measurements. Holding recent observations and filtering reception histories determine how much temporal information reaches that estimator. Their value may change when the position estimates themselves are temporally filtered.

We study these choices in a cattle-tracking deployment with four BLE gateways covering an analysis region of $4{,}302.4\,\mathrm{m}^2$. This deployment provides a field setting for comparing temporal input representations under intermittent reception.

Across the analysis population, an animal is heard by approximately 0.99 gateways per second, and all four gateways receive simultaneously in only 0.0096\% of animal-seconds. The resulting input is chronically incomplete. This motivates separating the benefit of retaining a short history from the additional effects of hold duration, input filtering, and downstream temporal memory.

A useful comparison must also distinguish information available online from observations that arrive later. We therefore evaluate causal input rules alongside explicitly offline references that can use future receptions.

We investigate temporal input handling using a random forest (RF) as the primary single-frame probe and $k$-nearest neighbours (kNN) as a secondary check. The study makes three empirical contributions:
\begin{enumerate}
\item We quantify the effects of short-history retention, longer holding, and input filtering using validation-selected parameters and paired uncertainty estimates.
\item We use representation ablations, direction controls, and specific offline and elapsed-time state-space references to delimit the interpretation of the filtering benefit.
\item We evaluate whether these comparisons persist in an ungated per-second pipeline with output temporal filtering, and assess sensitivity across held-out date blocks within the same record.
\end{enumerate}

The contribution is an empirical characterization of temporal input choices in this deployment. We first compare representations under the historical input-screening protocol, then examine their value in a per-second pipeline whose reception inputs are independent of GPS screening.

\section{Related Work}
BLE fingerprinting provides a foundation for learning-based localization. A study of indoor BLE positioning examines beacon density, transmission settings, and fast fading~\cite{faragher2015location}. In livestock monitoring, a barn localization system aggregates received measurements in non-overlapping 30-second windows and combines RSSI-derived distance rings with total signal strength~\cite{nikodem2021barn}. These studies motivate examining how observations are accumulated before localization. Our focus is the temporal availability and representation of observations in a sparse outdoor deployment; accuracy figures from different sites are not treated as directly comparable benchmarks.

Radio-map imputation can exploit temporal information in survey data. A study of sparse indoor radio maps distinguishes missing-at-random and missing-not-at-random RSSIs, then jointly imputes missing RSSIs and reference positions using a bidirectional sequence model with time-lag and attention mechanisms~\cite{li2023imputation}. Its offline map-construction setting differs from forming an input at prediction time using only available receptions. A separate study detects missing access points and rebuilds an indoor Wi-Fi localization model to accommodate infrastructure changes~\cite{gorak2018missingap}. Our comparison instead concerns intermittent reception histories from the deployed gateways. These distinctions concern the task and available information; sparsity alone does not exclude temporal or learned imputation methods.

Temporal state-space models offer a further way to handle intermittent observations. Prior work develops Kalman filtering and smoothing formulations of temporal Gaussian-process regression, with exact finite-dimensional representations for half-integer Mat\'ern kernels and approximations for the squared-exponential kernel~\cite{hartikainen2010temporal}. Motivated by elapsed-time state propagation, we evaluate random-walk and mean-reverting scalar controls with explicitly specified initialization. A study of leakage in machine-learning-based science also motivates separating parameter selection, scoring, and the information permitted to update an online state~\cite{kapoor2023leakage}.

\section{Deployment and Data}
\begin{figure}[t]
\centering
\includegraphics[width=\columnwidth]{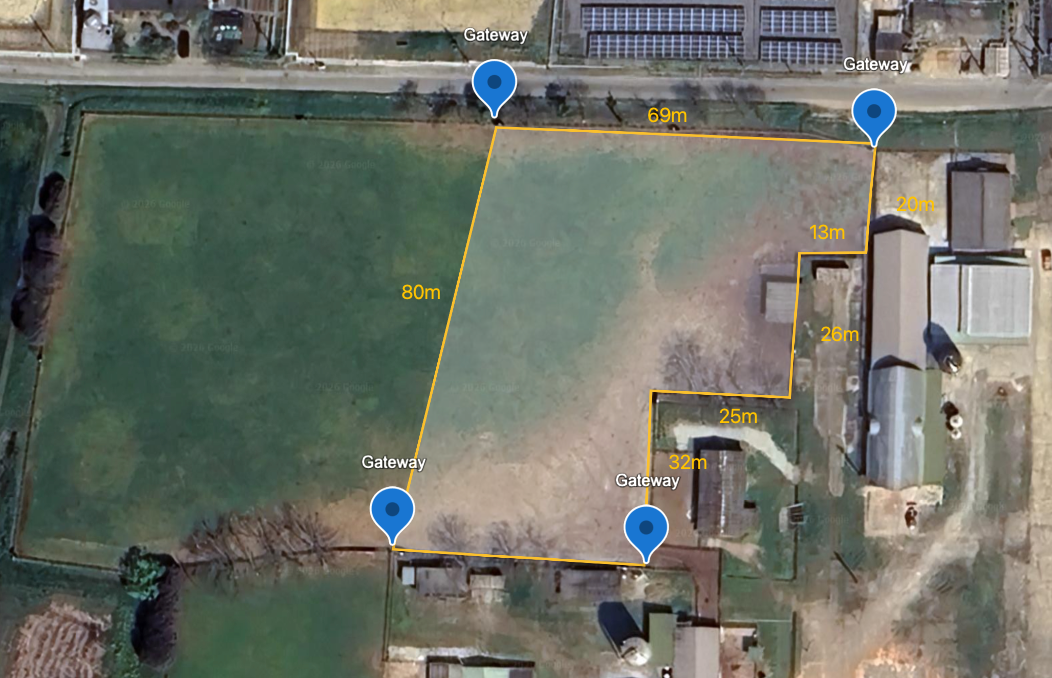}
\caption{The deployment, with the four gateway positions marked~\cite{google_earth_2024}. Quantitative analyses use the convex hull of these four positions, with an area of $4{,}302.4\,\mathrm{m}^2$.}
\label{fig:site}
\end{figure}

The site is an experimental cattle pasture operated by Kobe University in Hyogo, Japan (Fig.~\ref{fig:site}).
Four BLE gateways are mounted on $4\,\mathrm{m}$ poles at the pasture boundary with omnidirectional antennas, each scanning continuously for advertising packets from collar tags and logging RSSI.
Every collar carries a BLE beacon, a GPS module, and a nine-axis inertial measurement unit.
GPS positions provide training labels and reference positions for evaluation; the inertial measurements are not inputs to the probes evaluated here.
Quantitative results are computed over the convex hull of the four gateways, $4{,}302.4\,\mathrm{m}^2$; mean nearest-neighbour gateway separation is $54.2\,\mathrm{m}$.

The deployment has an anchor density of $0.93$ gateways per $1000\,\mathrm{m}^2$.
The ten-day record is organized on a one-second analysis grid; this grid does not specify the BLE advertising interval.
Two collars were excluded using a reception-coverage criterion independent of positioning error, leaving sixteen animals and $3{,}687{,}672$ eligible animal-second records.
Historical preprocessing also used GPS-based validity and analysis-region conditions to screen reception inputs.
Section~\ref{sec:protocol} distinguishes this gated protocol from the ungated evaluation, in which GPS supplies training labels and identifies scoring rows but does not determine which receptions enter the input stream.

\begin{table}[t]
\caption{Deployment geometry and reception statistics over the full analysis population. Temporal statistics refer to the one-second analysis grid.}
\label{tab:deploy}
\centering
\small
\begin{tabular}{@{}llr@{}}
\toprule
Axis & Quantity & Value \\
\midrule
Spatial  & Analysis region              & $4{,}302.4\,\mathrm{m}^2$ \\
         & Gateways                     & $4$ \\
         & Anchor density               & $0.93$ / $1000\,\mathrm{m}^2$ \\
         & Mean NN separation           & $54.2\,\mathrm{m}$ \\
\midrule
Temporal & Analysis grid                & $1\,\mathrm{Hz}$, $10$ days \\
         & Pr[a given gateway is heard  & \\
         & \quad in a given second]     & $0.247$ \\
         & \quad range over the four    & $0.218$--$0.266$ \\
         & Mean audible gateways / s    & $0.99$ \\
         & Seconds with no reception    & $29.6\%$ \\
         & Observation age, median       & $2\,\mathrm{s}$ \\
         & \quad 90th percentile         & $48\,\mathrm{s}$ \\
         & Complete 4-channel vectors   & $0.0096\%$ \\
\bottomrule
\end{tabular}
\end{table}

Table~\ref{tab:deploy} describes both deployment geometry and temporal reception sparsity over the full analysis population, rather than the fixed test subsample.
On average, approximately one gateway receives an animal's signal in a given second, and simultaneous reception on all four channels is rare.
This temporal sparsity motivates examining how recent observations are retained and filtered before they are supplied to a localization estimator.

\section{Evaluation Protocol}
\subsection{Observation model and gap-handling rules}
Let $r_i(t)$ denote the RSSI reported by gateway $i\in\{1,\dots,4\}$ at second $t$.
A gateway reports only when it receives an advertising packet, so $r_i(t)$ is defined on a sparse subset of seconds.
\begin{equation}
\tau_i(t)=\max\{\,u\le t:\ r_i(u)\ \text{is defined}\,\},\quad
a_i(t)=t-\tau_i(t),
\label{eq:age}
\end{equation}
where $\tau_i(t)$ is the time of the most recent reception on channel $i$ and $a_i(t)$ is the age of that reception at time $t$.
An estimator consumes a fixed-length vector $\mathbf{x}(t)=[x_1(t),\dots,x_4(t)]^{\mathsf T}$, one entry per gateway. Max-hold is
\begin{equation}
x_i(t)=
\begin{cases}
r_i(\tau_i(t)), & a_i(t)\le W,\\[2pt]
s, & \text{otherwise.}
\end{cases}
\label{eq:maxhold}
\end{equation}
Here $W$ is the maximum age of a retained observation and $s=-110\,\mathrm{dBm}$ is the sentinel. Before the first reception on a channel, we set $a_i(t)=\infty$ and use the sentinel explicitly, including when $W=\infty$. We denote each configuration by its hold window, such as W4 or W120. Setting $W=0$ retains only current observations, and $W\to\infty$ holds indefinitely after a reception becomes available.

The input-filtering configuration $f_\alpha$ replaces the most recent raw value with a scalar Kalman estimate updated at reception events and held between them without an age limit. The parameter $\alpha$ identifies the steady-state gain; the gain during initialization need not equal $\alpha$. For reception index $k$, the measurement update has the form
\begin{equation}
\widehat h_{i,k}=\widehat h^-_{i,k}
 +K_{i,k}\bigl(r_{i,k}-\widehat h^-_{i,k}\bigr),
\label{eq:input_update}
\end{equation}
where $\widehat h^-_{i,k}$ is the prior estimate and $K_{i,k}$ is the Kalman gain. For $0<\alpha<1$, we use a random-walk state model with observation-noise variance $R=1$ and process-noise variance $Q=\alpha^2/(1-\alpha)$ per reception transition. Writing $P$ for the posterior estimation-error variance, subsequent receptions use $P^-=P+Q$, $K=P^-/(P^-+R)$, and $P=(1-K)P^-$; the gain converges to $\alpha$, with smaller $\alpha$ placing less steady-state weight on the current observation. At the first reception, the prior mean equals that observation and the prior variance is 1; the measurement update adds no process noise and leaves the mean unchanged with posterior variance $1/2$. Between receptions, neither mean nor variance is propagated; inputs are held without an age limit, use $-110\,\mathrm{dBm}$ before the first reception, and reduce to unbounded raw holding at $\alpha=1$.

The corresponding $s_\alpha$ configuration applies Rauch--Tung--Striebel (RTS) smoothing to the same reception-indexed model within each animal-day, then holds the smoothed estimates between receptions; it is an offline reference. Processing is separate by animal, gateway, and day, with states reset at daily boundaries.

A rule is online when its input at time $t$ depends only on receptions available by $t$. Backward filling and RTS smoothing use future receptions and are offline controls. The main comparisons use holding, input filtering, and RTS. Representation controls additionally replace values with filtered estimates only at observed or only at missing channel-seconds.

\subsection{Protocol}
\label{sec:protocol}
The main evaluation trains on days 1--8 and scores 50,000 fixed eligible rows from days 9--10, using seeds 42, 43, and 44. Parameter selection first trains on days 1--6 and evaluates 50,000 fixed validation rows from days 7--8. The selected settings are then used in the main evaluation. Both probes select W120; RF selects $\alpha=0.1$ for filtering and RTS, and kNN selects 0.05 for each. Days 9--10 had been inspected in earlier analyses, so they are historical test days rather than an untouched test set. Test-grid optima are reported only descriptively.

The hold windows evaluated are 0, 4, 8, 16, 32, 64, 120, 240, 480, and 960 s, together with unbounded holding. The input-filtering values of $\alpha$ are 0.005, 0.01, 0.02, 0.03, 0.05, 0.1, 0.2, 0.3, 0.5, 0.7, and 1; the RTS grid excludes 1. Selection minimizes mean validation error across seeds.

We distinguish historical gated and ungated inputs. Gated preprocessing retains receptions only where the GPS-validity, finite-coordinate, and analysis-region conditions hold. Ungated preprocessing retains all receptions, while GPS supplies training labels and identifies scoring rows. Only the latter input flow can be used without GPS deciding when online states update. Parameters selected under the historical protocol are retained in the ungated input comparison; the output-filter coefficient is selected separately as described in Sec.~\ref{sec:memory}.

\subsection{Probes}
RF is the primary probe and kNN is a secondary check. Both predict two-dimensional metric coordinates from the four-channel representation at a single time. RF uses 100 trees and a training subsample of 150,000 rows per seed; kNN uses $k=4$ and 30,000 rows. Features retain their dBm scale without standardization. Each representation is used to retrain the probe under the same prescribed model settings and training budget. The revision experiments use scikit-learn 1.5.1.

Single-frame probes receive temporal information through the input representation. The separate output-filtering evaluation tests how the comparisons change when downstream processing also retains a history of predictions.

\subsection{Error and uncertainty}
We report mean Euclidean localization error in metres. For paired contrasts, the three seed errors are averaged for each scoring row before differences are formed on identical rows. A positive contrast $e(A)-e(B)$ indicates lower error for B. The main 95\% confidence intervals use 2,000 animal-cluster bootstrap resamples of sixteen animals. Each sampled animal contributes all its evaluated rows, and the pooled mean is row-weighted. These intervals are conditional on the evaluated dates and fitted models; they do not capture all selection or training uncertainty.

Animal-equal-weight estimates and animal-by-day block intervals provide sensitivity checks. The latter resample whole animal-day blocks, not fixed-length moving windows. Seed standard deviations describe fitting variability. Error medians and 90th percentiles are computed within each seed and then averaged across seeds.

\section{Holding and Input Filtering}
\label{sec:results}
\subsection{Short histories and hold duration}
\label{sec:hold}
Under the historical gated protocol, RF error decreases from 27.632 m for W0 to 19.557 m for W4. The paired reduction is 8.075 m [6.787, 9.454]. Extending the hold from W4 to validation-selected W120 provides a smaller reduction of 0.271 m [0.174, 0.380]. Table~\ref{tab:primary} reports corresponding comparisons for both probes.

Across the RF hold grid, the error range for $W\geq4$ s is 0.310 m. The W0-to-W4 reduction accounts for 96.75\% of the reduction to validation-selected W120 in this record. Thus, retaining a short history provides most of the observed gain, with a smaller additional benefit from the longer selected hold.

\begin{table}[t]
\caption{Historical gated protocol: paired error reductions (m), with 95\% animal-cluster intervals, on the same 50,000 test rows. Positive A$-$B favours B. Parameters are independently validation-selected: RF $\alpha=0.1$, kNN $\alpha=0.05$.}
\label{tab:primary}
\centering\small
\begin{tabular}{@{}lcc@{}}
\toprule
Contrast & RF & kNN \\
\midrule
W0$-$W4 & 8.075 & 8.149 \\
 & [6.787, 9.454] & [6.901, 9.477] \\
W4$-$W120 & 0.271 & 0.243 \\
 & [0.174, 0.380] & [0.054, 0.420] \\
W120$-f_\alpha$ & 0.736 & 1.466 \\
 & [0.224, 1.267] & [0.935, 2.017] \\
W$\infty-f_\alpha$ & 0.969 & 1.552 \\
 & [0.417, 1.538] & [1.061, 2.030] \\
$f_\alpha-s_\alpha$ & 0.227 & 0.336 \\
 & [0.025, 0.424] & [0.139, 0.561] \\
\bottomrule
\end{tabular}
\end{table}

\subsection{Validation-selected input filtering}
\label{sec:alpha}
Relative to W120, validation-selected filtering reduces RF error by 0.736 m [0.224, 1.267] and kNN error by 1.466 m [0.935, 2.017]. The RF mean errors are $19.29\pm0.04$ m for W120 and $18.55\pm0.03$ m for $f_{0.1}$, where the deviations are across seeds. Their medians are 17.25 and 15.91 m, and their 90th percentiles are 37.29 and 36.78 m. Absolute errors therefore remain substantial despite the improvement.

Stronger filtering is not uniformly beneficial. Changing RF from $f_{0.1}$ to $f_{0.005}$ increases error by 2.689 m [1.787, 3.494], showing the cost of stronger filtering in this configuration.

\subsection{Representation controls}
\begin{figure}[t]
\centering
\includegraphics[width=\columnwidth]{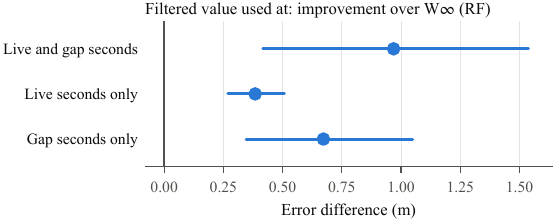}
\caption{RF error reductions relative to W$\infty$ under historical gated inputs ($\alpha=0.1$). The filtered estimate replaces raw values at observed channel-seconds (live-only), missing channel-seconds (gap-only), or both (full). Each configuration is retrained and scored on the same 50,000 rows. Bars are 95\% animal-cluster intervals.}
\label{fig:representation}
\end{figure}

We retrain RF with filtered values substituted at observed channel-seconds only, missing channel-seconds only, or both (Fig.~\ref{fig:representation}). Relative to W$\infty$, the respective error reductions are 0.384, 0.672, and 0.969 m, with positive animal-cluster intervals in all three cases. All configurations are scored on the same rows; observed and missing refer to where channel values are replaced, not to disjoint evaluation subsets. The full-minus-live-minus-gap interaction is $-0.088$ m [$-0.398$, 0.163]. These comparisons do not form an additive decomposition of a physical mechanism.

For RF, backward filling minus W$\infty$ is 0.085 m [$-0.006$, 0.170] under row weighting, but 0.112 m [0.022, 0.201] under animal-equal weighting. In 18,606 rows where every missing channel has past and future reception distances differing by at most 2 s, the difference is $-0.041$ m [$-0.150$, 0.067]. No clear directional difference is detected in that subset. This matching restricts the past--future temporal-distance imbalance while leaving motion direction and detection geometry uncontrolled.

\subsection{Offline and elapsed-time references}
The RF difference between $f_{0.1}$ and $s_{0.1}$ is 0.227 m [0.025, 0.424]. This quantifies the difference to one validation-selected RTS reference; it is not a bound on the improvement available to all offline methods.

We additionally evaluate elapsed-time random-walk and mean-reverting scalar filters under the historical gated protocol. For the random walk, uncertainty propagates as $P^-=P+q\Delta t$, with the last filtered mean retained between receptions. For mean reversion, $a=\exp(-\Delta t/\tau)$, $x^-=\mu+a(x-\mu)$, and $P^-=a^2P+(q\tau/2)(1-a^2)$; the mean is propagated towards the training-derived $\mu$ between receptions. Both use unit observation noise and initial variance and initialize the state with the first reception.

The process-noise values evaluated are $10^{-4}$, $3\times10^{-4}$, $10^{-3}$, $3\times10^{-3}$, $10^{-2}$, and $3\times10^{-2}$ per second, relative to unit observation noise. Reversion times are 60, 300, and 1800 s. Selection uses the same validation dates. The RF-selected mean-reverting configuration uses $q=10^{-3}$ per second and $\tau=1800\,\mathrm{s}$. This reversion time is at the upper grid boundary, limiting conclusions about that model family.

RF validation selects the random walk with $q=3\times10^{-4}$ per second over the selected mean-reverting configuration. The error difference $f_{0.1}$ minus the selected random walk is 0.193 m [$-0.076$, 0.474]; the corresponding mean-reverting comparison is 0.196 m [$-0.051$, 0.456]. Neither interval establishes a clear difference or equivalence. For kNN using the same RF-selected configurations, the random-walk comparison is 0.359 m [0.139, 0.586]. The findings therefore depend on the probe and the evaluated configurations.

\section{Output Memory and Sensitivity}
\label{sec:memory}
\subsection{Per-second output filtering}
Using ungated reception inputs, we generate RF positions at every second, including seconds without a valid GPS scoring label. A causal output filter updates
\begin{equation}
\widetilde{\mathbf p}(t)=\beta\widehat{\mathbf p}(t)
+(1-\beta)\widetilde{\mathbf p}(t-1).
\label{eq:output_filter}
\end{equation}
The first prediction initializes the state, which resets at each animal-day boundary. GPS does not determine state-update times. Input parameters retain their historical validation choices. Each rule independently selects $\beta$ on days 7--8 using models trained on days 1--6. The values evaluated are 1, 0.5, 0.2, 0.1, 0.05, 0.02, 0.01, 0.005, 0.002, and 0.001. Table~\ref{tab:stream_absolute} reports the selected coefficients and absolute errors.

\begin{table}[t]
\caption{RF mean Euclidean errors (m) with ungated reception inputs, averaged over three seeds on the same 50,000 historical test rows. Input parameters retain their historical validation choices; each output coefficient is separately validation-selected. Selected paired contrasts and their uncertainty are shown in Fig.~\ref{fig:memory}.}
\label{tab:stream_absolute}
\centering\small
\begin{tabular}{@{}lrrr@{}}
\toprule
Input rule & Single frame & With output filter & $\beta$ \\
\midrule
W0 & 27.632 & 26.095 & 0.1 \\
W4 & 19.540 & 18.759 & 0.1 \\
W120 & 19.307 & 18.243 & 0.02 \\
W$\infty$ & 19.518 & 18.472 & 0.005 \\
$f_{0.1}$ & 18.498 & 18.121 & 0.005 \\
\midrule
\multicolumn{4}{@{}l}{\emph{Offline input reference}} \\
$s_{0.1}$ & 18.208 & 17.631 & 0.005 \\
\bottomrule
\end{tabular}
\end{table}

\begin{figure}[t]
\centering
\includegraphics[width=\columnwidth]{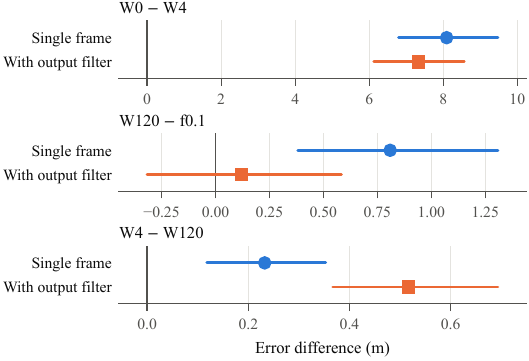}
\caption{Paired RF error differences under ungated inputs, before and after output filtering. Each rule selects its output coefficient on the validation days. Bars are 95\% animal-cluster bootstrap intervals on identical historical test rows. Positive values favour the second configuration in each contrast. Horizontal scales differ between panels.}
\label{fig:memory}
\end{figure}

Fig.~\ref{fig:memory} compares configurations within the ungated protocol. The W0-to-W4 reduction remains 7.336 m [6.127, 8.568] after output filtering. The additional benefit of $f_{0.1}$ over W120 is 0.810 m [0.381, 1.309] without output filtering and 0.121 m [$-0.317$, 0.584] with it. The latter interval leaves the sign uncertain and does not establish zero benefit. No paired interval for the difference between these two benefits was estimated.

The W4-minus-W120 difference is 0.517 m [0.367, 0.693] in the cascade, compared with 0.232 m [0.118, 0.353] without output filtering. These comparisons use independently selected output coefficients, not a fixed-$\beta$ factorial design. Cascaded W120 and $f_{0.1}$ have mean errors of 18.243 and 18.121 m, respectively. Cascaded $s_{0.1}$ remains an offline-input reference even though its output filter is causal.

\subsection{Input screening and heterogeneity}
For single-frame RF, ungated-minus-gated error is $-0.017$ m [$-0.027$, $-0.007$] at W4 and 0.022 m [$-0.008$, 0.059] at W120. The corresponding RF intervals across the hold grid lie within the prespecified engineering band of $\pm0.2$ m. This band is an operational convention, not a statistically derived threshold. The conclusion is not uniform across probes: kNN at W120 changes by 0.138 m [0.030, 0.296], and some longer holds have wider intervals outside the band.

In the main gated RF evaluation, the W$\infty$-to-$f_{0.1}$ benefit is approximately 1.86 m on day 9 and 0.35 m on day 10. The global setting increases error for three animals; only two of these have an exploratory test-grid optimum at no filtering. These differences limit interpretation of the pooled benefit as a uniform effect.

\subsection{Held-out date blocks}
Five folds each hold out two consecutive days and train on the remaining eight, with 50,000 scoring rows per fold. Inputs use the historical gated protocol and parameters remain fixed at the original RF-selected values. The first four folds train on dates later than their test dates, and days 7--8 participated in the original parameter selection. These are within-record sensitivity analyses, not nested selection or prospective validation.

\begin{figure}[t]
\centering
\includegraphics[width=\columnwidth]{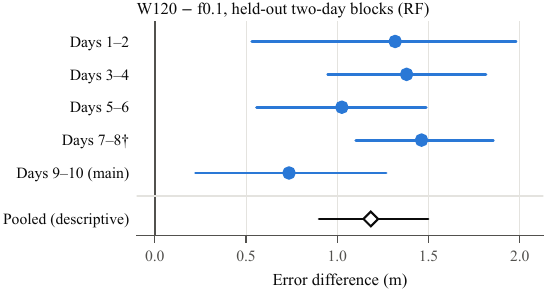}
\caption{RF W120-minus-$f_{0.1}$ differences across held-out two-day blocks under historical gated inputs. Bars are 95\% animal-cluster intervals. Parameters are fixed across folds. The marked days previously served for parameter selection; the first four folds train on later dates. Pooling keeps each animal's records across folds in one cluster and is descriptive.}
\label{fig:days}
\end{figure}

RF W0-to-W4 differences are positive in every fold, ranging from 5.770 to 10.067 m. The benefit of $f_{0.1}$ over W120 also has a positive interval in each fold (Fig.~\ref{fig:days}); its descriptive pooled value is 1.184 m [0.901, 1.497]. For kNN using the shared RF configuration, W120 minus $f_{0.1}$ is $-1.024$ m [$-4.988$, 1.580] in the first fold and 0.696 m [$-0.315$, 1.430] when pooled. Overlapping training sets and prior parameter selection preclude treating the folds as independent confirmations of future performance.

\section{Discussion and Limitations}
Retaining recent RSSI history produces the largest observed improvement over instantaneous inputs in this deployment. Input filtering adds a smaller single-frame benefit, while its additional benefit over W120 is uncertain after the independently selected output filters are applied. The W4-to-W120 improvement remains clear in that cascade. These findings support comparing hold duration and input filtering within the intended prediction pipeline, including its downstream temporal processing.

For this deployment, a useful starting point is therefore a recent-history representation followed by validation of both the hold window and any additional filtering. Absolute errors remain substantial: the ungated RF cascade gives mean errors of 18.243 m with W120 and 18.121 m with $f_{0.1}$. Their suitability depends on the tracking application's spatial requirements. The selected windows and gains are specific to the evaluated data and models; agreement between RF and kNN does not establish estimator independence.

The controls do not identify a unique physical mechanism. Filtering modifies both observed and carried-forward values, and each representation induces a separately fitted predictor. Reception can depend on motion, position, and propagation conditions. Past and future samples need not be exchangeable even after temporal-distance matching; neither these comparisons nor the live/gap ablations exclude reconstruction effects. The offline and elapsed-time controls cover specific configurations, leaving other temporal and learned imputation methods open for comparison.

All measurements come from one site, sixteen animals, and one ten-day record. The main test dates were previously explored, and additional folds do not remove that history. Animal-cluster intervals do not establish robustness across sites, seasons, or all model-selection decisions. The streaming experiment resets at daily boundaries and tests one form of output memory. Learned sequence models, their training-budget sensitivity, and a quantitative link from RSSI noise statistics to localization gains remain outside the evaluated scope.

\section{Conclusion}
In this sparse outdoor BLE deployment, retaining a short RSSI history produces the largest observed improvement over instantaneous sentinel inputs. Validation-selected input filtering provides an additional single-frame benefit, while its extra benefit over W120 remains uncertain in the tested output-filtering cascade. The practical implication is to evaluate input history and filtering together with downstream temporal processing, using validation-selected configurations and paired comparisons in the target deployment.

\section*{Acknowledgment}
The authors would like to thank ChatGPT for its assistance in English language editing, grammatical correction, and readability enhancement during the preparation of this manuscript. 
\bibliographystyle{IEEEtran}
\bibliography{refs_ws16}

\end{document}